\documentclass[a4paper,10pt,twoside]{cpc-hepnp}

\usepackage{multicol}
\usepackage{graphicx}
\usepackage{booktabs}
\usepackage{amssymb,bm,mathrsfs,bbm,amscd}
\usepackage[tbtags]{amsmath}
\usepackage{lastpage}

\begin{document}

\fancyhead[co]{\footnotesize Lei Chang et al: Exploring the light-quark interaction}

\footnotetext[0]{Received 2 July 2009}

\title{Exploring the light-quark interaction\thanks{%
This work was supported by: 
the National Natural Science Foundation of China, contract no.\ 10705002; and
the Department of Energy, Office of Nuclear Physics, contract nos.\ DE-FG03-97ER4014 and DE-AC02-06CH11357.}}

\author{%
      Lei CHANG$^{1}$%
\quad Ian C.\ CLO\"ET$^2$%
\quad Bruno El-BENNICH$^3$%
\quad Thomas KL\"AHN$^3$%
\quad Craig D. ROBERTS$^{3,4;1)}$\email{cdroberts@anl.gov}%
}
\maketitle

\address{%
1~(Institute of Applied Physics and Computational Mathematics, Beijing 100094, China)\\
2~(Department of Physics, University of Washington, Seattle WA 98195, USA)\\
3~(Physics Division, Argonne National Laboratory, Argonne,
Illinois 60439, USA)\\
4~(Department of Physics, Peking University, Beijing 100871, China)\\
}

\begin{abstract}
Two basic motivations for an upgraded JLab facility are the needs: to determine the essential nature of light-quark confinement and dynamical chiral symmetry breaking (DCSB); and to understand nucleon structure and spectroscopy in terms of QCD's elementary degrees of freedom.  During the next ten years a programme of experiment and theory will be conducted that can address these questions.  We present a Dyson-Schwinger equation perspective on this effort with numerous illustrations, amongst them: an interpretation of string-breaking; a symmetry-preserving truncation for mesons; the nucleon's strangeness $\sigma$-term; and the neutron's charge distribution.
\end{abstract}



\begin{multicols}{2}

\section{Confinement and dynamical chiral symmetry breaking}
Understanding the spectrum of hadrons with masses less than 2\,GeV and their interactions is an essential step toward revealing the essence of light-quark confinement and dynamical chiral symmetry breaking (DCSB), and describing hadron structure in terms of QCD's elementary degrees of freedom.  These are basic questions, which define a frontier of contemporary hadron physics.

In connection with confinement it is important to appreciate that the static potential measured in numerical simulations of quenched lattice-regularised QCD is not related in any known way to the question of light-quark confinement.  It is a basic feature of QCD that light-quark creation and annihilation effects are essentially nonperturbative. Therefore it is impossible in principle to compute a potential between two light quarks.

It is known\cite{Bali:2005fu} that in the presence of two dynamical flavours of quark, each with a current-quark mass $\sim m_s$; i.e., typical of the $s$-quark, string breaking is a nonlocal and instantaneous process, which occurs when the static quark and antiquark are separated by $\approx 1.25$fm.  There is therefore a critical energy connected with the string; viz., $E_c \approx 1.25\,$GeV.  

It is noteworthy and instructive that $E_c \simeq M_S+M_{\bar S}$, where $M_S$ and $M_{\bar S}$ are, respectively, constituent-quark masses associated with the lightest quark and antiquark in the system; namely, the $s$-quark in this instance.  Our observation suggests an intuitive understanding of string breaking; namely, the flux tube collapses instantly and entirely when the energy it contains exceeds that required to produce the lightest constituent quark-antiquark pair, and the distorted and distressed upsilon-like state switches instantly to a pair of localised heavy-light mesons.


Our estimate of $M_S=M_{\bar S}$ is based on extensive experience with QCD's Dyson-Schwinger equations (DSEs).\cite{Roberts:1994dr,Maris:2003vk,Roberts:2007jh,Roberts:2007ji}  Typically, $m_s \simeq 25\,m_u$ and $M_S \simeq M_U+0.15\,$GeV$\simeq 0.55\,$GeV.  The phenomenon underlying this magnification of the current-quark mass is DCSB, which can be understood via the renormalised gap equation: 
\begin{eqnarray}
\nonumber 
\lefteqn{S(p)^{-1} =  Z_2 \,(i\gamma\cdot p + m^{\rm bm})} \\
&+&  Z_1 \int^\Lambda_q\! g^2 D_{\mu\nu}(p-q) \frac{\lambda^a}{2}\gamma_\mu S(q) \Gamma^a_\nu(q,p) , \label{gendse}
\end{eqnarray}
where $\int^\Lambda_q$ indicates a Poincar\'e invariant regularisa- %
\begin{center}
\includegraphics[clip,width=0.4\textwidth]{Mp2Jlab.eps}
\figcaption{\label{gluoncloud} Dressed-quark mass function, $M(p)$: solid curves -- DSE results,\protect\cite{Bhagwat:2003vw,Bhagwat:2006tu} ``data'' -- numerical simulations of unquenched lattice-QCD.\protect\cite{Bowman:2005vx} In this figure one observes the current-quark of perturbative QCD evolving into a constituent-quark as its momentum becomes smaller.  The constituent-quark mass arises from a cloud of low-momentum gluons attaching themselves to the current-quark.  This is dynamical chiral symmetry breaking: an essentially nonperturbative effect that generates a quark mass \emph{from nothing}; namely, it occurs even in the chiral limit.}
\end{center}
tion of the integral, with $\Lambda$ the regularisation mass-scale, $D_{\mu\nu}$ is the renormalised dressed-gluon propagator, $\Gamma_\nu$ is the renormalised dressed-quark-gluon vertex, and $m^{\rm bm}$ is the quark's $\Lambda$-dependent bare current-mass.  The vertex and quark wave-function renormalisation constants, $Z_{1,2}(\zeta^2,\Lambda^2)$, depend on the gauge parameter.  The solution to Eq.\,(\ref{gendse}) has the form
\begin{eqnarray} 
 S(p) & =&  
\frac{Z(p^2,\zeta^2)}{i\gamma\cdot p + M(p^2)}\,
%
\label{Sgeneral}
\end{eqnarray} 
and it is important that the mass function, $M(p^2)=B(p^2,\zeta^2)/A(p^2,\zeta^2)$ is independent of the renormalisation point, $\zeta$.  The form this function takes in QCD is depicted in Fig.\,\ref{gluoncloud}.

The behaviour of the dressed-quark mass function is one of the most remarkable features of the Standard Model.  In perturbation theory it is impossible in the chiral limit to obtain $M(p^2)\neq 0$: the generation of mass \emph{from nothing} is an essentially nonperturbative phenomenon.  On the other hand, it is a longstanding prediction of nonperturbative DSE studies that DCSB will occur so long as the integrated infrared strength possessed by the gap equation's kernel exceeds some critical value.\cite{Roberts:1994dr}  There are strong indications that this condition is satisfied in QCD.\cite{Bhagwat:2003vw,Bhagwat:2006tu,Bowman:2005vx}  

It follows that the quark-parton of QCD acquires a momentum-dependent mass, which at infrared momenta is roughly $100$-times larger than the light-quark current-mass.  This effect owes primarily to a dense cloud of gluons that clothes a low-momentum quark.  It means that the Higgs mechanism is largely irrelevant to the bulk of normal matter in the universe.  Instead, the single most important mass generating mechanism for light-quark hadrons is the strong interaction effect of DCSB; e.g., one may identify it as being responsible for 98\% of a proton's mass. 

Confinement can be connected with the analytic properties of QCD's Schwinger functions.\cite{Krein:1990sf,Roberts:1994dr,Roberts:2007ji,Roberts:2007jh}  Indeed, the presence of an inflexion point in the DSE prediction for the dressed-quark mass function, which lattice simulations may be argued to confirm, signals confinement of the dressed-quark.\cite{Roberts:2007jh}  Kindred behaviour is observed in the gluon and ghost self-energies.\cite{Bowman:2007du,Cucchieri:2008fc}

From this standpoint the question of light-quark confinement can be translated into the challenge of charting the infrared behavior of QCD's \emph{universal} $\beta$-function. (Although this function may depend on the scheme chosen to renormalise the theory, it is unique within a given scheme.)
This is a well-posed problem whose solution is an elemental goal of modern hadron physics and which can be addressed in any framework enabling the nonperturbative evaluation of renormalisation constants. 

Through the gap and Bethe-Salpeter equations (BSEs) the pointwise behaviour of the $\beta$-function determines the nature  of chiral symmetry breaking; e.g., the evolution in Fig.\,\ref{gluoncloud}.  Moreover, the fact that DSEs connect the $\beta$-function to experimental observables entails that comparison between computations and observations of hadron properties can be used to chart the $\beta$-function's long-range behaviour. 

\section{DSE truncations:\\ preserving symmetry}
In order to realise this goal a nonperturbative symmetry-preserving DSE truncation is necessary.  Steady quantitative progress continues with a scheme that is systematically improvable.\cite{Munczek:1994zz,Bender:1996bb} Indeed, its mere existence has enabled the proof of exact nonperturbative results in QCD.  Amongst them are veracious statements about the $\eta$-$\eta^\prime$ complex and $\pi^0$-$\eta$-$\eta^\prime$ mixing, with predictions of $\theta_{\eta \eta^\prime} = -15^\circ$ and $\theta_{\pi^0 \eta} = 1^\circ$.\cite{Bhagwat:2007ha}  Only studies that are demonstrably consistent with the results proved therein can be considered seriously.  

It is also true that significant qualitative advances can be made with symmetry-preserving kernel \emph{Ans\"atze} that express important additional nonperturbative effects, which are difficult to capture in any finite sum of contributions.\cite{Chang:2009zb}  In order to elucidate we consider the example of pseudoscalar and axial-vector mesons, which appear as poles in the inhomogeneous BSE for the axial-vector vertex, $\Gamma_{5\mu}^{fg}$.  An exact form of that equation is ($q_\pm = q\pm P/2$, etc.)
\begin{eqnarray}
\nonumber
\lefteqn{\Gamma_{5\mu}^{fg}(k;P) = Z_2 \gamma_5\gamma_\mu - \int_q g^2D_{\alpha\beta}(k-q)\, }\\
\nonumber
&& \rule{-2em}{0ex} \times \frac{\lambda^a}{2}\,\gamma_\alpha S_f(q_+) \Gamma_{5\mu}^{fg}(q;P) S_g(q_-) \frac{\lambda^a}{2}\,\Gamma_\beta^g(q_-,k_-) \\
&& \rule{-2em}{0ex} + \int_q g^2D_{\alpha\beta}(k-q)\, \frac{\lambda^a}{2}\,\gamma_\alpha S_f(q_+) \frac{\lambda^a}{2} \Lambda_{5\mu\beta}^{fg}(k,q;P), \label{genbse}
\end{eqnarray}
where $\Lambda_{5\mu\beta}^{fg}$ is a 4-point Schwinger function that is completely defined via the quark self-energy.\cite{Munczek:1994zz,Bender:1996bb}  The pseudoscalar vertex, $\Gamma_5^{fg}(k;P)$, satisfies an analogous equation and has the general form
\begin{eqnarray}
\nonumber
\lefteqn{i\Gamma_{5}^{fg}(k;P) = \gamma_5 \left[ i E_5(k;P) + \gamma\cdot P F_5(k;P) \right.}\\
&& \left. + \gamma\cdot k \, G_5(k;P) + \sigma_{\mu\nu} k_\mu P_\nu H_5(k;P) \right].
\label{genG5}
\end{eqnarray}

In any dependable study of light-quark hadrons the solution of Eq.\,(\ref{genbse}) must satisfy the axial-vector Ward-Takahashi; viz., 
\begin{eqnarray}
\nonumber 
&& P_\mu \Gamma_{5\mu}^{fg}(k;P) + \, i\,[m_f(\zeta)+m_g(\zeta)] \,\Gamma_5^{fg}(k;P)\\
&=&S_f^{-1}(k_+) i \gamma_5 +  i \gamma_5 S_g^{-1}(k_-) \,,
\label{avwtim}
\end{eqnarray}
which expresses chiral symmetry and its breaking pattern.  The condition 
\begin{eqnarray}
\nonumber && P_\mu \Lambda_{5\mu\beta}^{fg}(k,q;P) + i [m_f(\zeta)+m_g(\zeta)] \Lambda_{5\beta}^{fg}(k,q;P)\\
& =& 
\Gamma_\beta^f(q_+,k_+) \, i\gamma_5+ i\gamma_5 \, \Gamma_\beta^g(q_-,k_-) \label{LavwtiGamma}  
\end{eqnarray}
where $\Lambda_{5\beta}^{fg}$ is the analogue of $\Lambda_{5\mu\beta}^{fg}$ in the pseudoscalar equation, is necessary and sufficient to ensure the Ward-Takahashi identity is satisfied.\cite{Chang:2009zb}  

Consider Eq.\,(\ref{LavwtiGamma}).  Rainbow-ladder is the lead\-ing-or\-der term in a systematic DSE truncation scheme.\cite{Munczek:1994zz,Bender:1996bb}  It corresponds to $\Gamma_\nu^f=\gamma_\nu$, in which case Eq.\,(\ref{LavwtiGamma}) is solved by $\Lambda_{5\mu\beta}^{fg}\equiv 0 \equiv \Lambda_{5\beta}^{fg}$.  This is the solution that indeed provides the rainbow-ladder forms of Eq.\,(\ref{genbse}).  Such consistency will be apparent in any valid systematic term-by-term improvement of the rainbow-ladder truncation.  

However, Eq.\,(\ref{LavwtiGamma}) is far more than merely a device for checking a truncation's consistency.  For, just as the vector Ward-Takahashi identity has long been used to build \emph{Ans\"atze} for the dressed-quark-photon vertex (e.g., Refs.\,\raisebox{-1.35ex}{\mbox{\Large \cite{Roberts:1994dr,Ball:1980ay,Kizilersu:2009kg}}}), Eq.\,(\ref{LavwtiGamma}) provides a tool for constructing a symmetry preserving kernel of the BSE that is matched to any reasonable \emph{Ansatz} for the dressed-quark-gluon vertex which appears in the gap equation.  With this powerful capacity Eq.\,(\ref{LavwtiGamma}) achieves a goal that has been sought ever since the Bethe-Salpeter equation was introduced.\cite{Salpeter:1951sz}  The symmetry-preserving kernel it can provide promises to enable the first reliable Poincar\'e invariant calculation of the spectrum of mesons with masses larger than 1\,GeV.

The utility of Eq.\,(\ref{LavwtiGamma}) can be illustrated through an application to ground state pseudoscalar and scalar mesons composed of equal-mass $u$- and $d$-quarks.  To this end, suppose that in Eq.\,(\ref{gendse}) one employs an \emph{Ansatz} for the quark-gluon vertex which satisfies
\begin{equation}
P_\mu i \Gamma_\mu^f(k_+,k_-) = {\cal B}(P^2)\left[ S_f^{-1}(k_+) - S_f^{-1}(k_-)\right]\,, \label{wtiAnsatz}
\end{equation}
with ${\cal B}$ flavour-independent.  (NB.\ While the true quark-gluon vertex does not satisfy this identity, owing to the form of the Slavnov-Taylor identity which it does satisfy, it is plausible that a solution of Eq.\,(\protect\ref{wtiAnsatz}) can provide a reasonable pointwise approximation to the true vertex.)  Given Eq.\,(\ref{wtiAnsatz}), then Eq.\,(\ref{LavwtiGamma}) entails ($l=q-k$)
\begin{equation}
i l_\beta \Lambda_{5\beta}^{fg}(k,q;P) =
{\cal B}(l)^2\left[ \Gamma_{5}^{fg}(q;P) - \Gamma_{5}^{fg}(k;P)\right], \label{L5beta}
\end{equation}
with an analogous equation for $P_\mu l_\beta i\Lambda_{5\mu\beta}^{fg}(k,q;P)$.  This identity can be solved to obtain
\begin{eqnarray}
\Lambda_{5\beta}^{fg}(k,q;P) & := & {\cal B}((k-q)^2)\, \gamma_5\,\overline{ \Lambda}_{\beta}^{fg}(k,q;P) \,, \label{AnsatzL5a}
\end{eqnarray}
with, using Eq.\,(\ref{genG5}), 
\begin{eqnarray}
\nonumber
\lefteqn{
\overline{ \Lambda}_{\beta}^{fg}(k,q;P) =  2 \ell_\beta \, [ i \Delta_{E_5}(q,k;P)+ \gamma\cdot P \Delta_{F_5}(q,k;P) ]}\\
\nonumber
&& +  \gamma_\beta \, \Sigma_{G_5}(q,k;P) +
2 \ell_\beta \,  \gamma\cdot\ell\, \Delta_{G_5}(q,k;P)  + [ \gamma_\beta,\gamma\cdot P]\\
&& \times \Sigma_{H_5}(q,k;P) + 2 \ell_\beta  [ \gamma\cdot\ell ,\gamma\cdot P]  \Delta_{H_5}(q,k;P) \,,
\label{AnsatzL5b}
\end{eqnarray}
where $\ell=(q+k)/2$, $\Sigma_{\Phi}(q,k;P) = [\Phi(q;P)+\Phi(k;P)]/2$ and $\Delta_{\Phi}(q,k;P) = [\Phi(q;P)-\Phi(k;P)]/[q^2-k^2]$.

Now, given any \emph{Ansatz} for the quark-gluon vertex that satisfies Eq.\,(\ref{wtiAnsatz}), then the pseudoscalar analogue of Eq.\,(\ref{genbse}) and Eqs.\,(\ref{gendse}), (\ref{AnsatzL5a}), (\ref{AnsatzL5b}) provide a symmetry-preserving closed system whose solution predicts the properties of pseudoscalar mesons.  
%
The relevant scalar meson equations are readily derived.  (NB.\ We are aware of the role played by resonant contributions to the kernel in the scalar channel \protect\cite{Holl:2005st} but they are not pertinent to this discussion.)
With these systems one can anticipate, elucidate and understand the impact on hadron properties of the rich nonperturbative structure expected of the fully-dressed quark-gluon vertex in QCD.

To proceed one need only specify the gap equation's kernel because the BSEs are completely defined therefrom.  To complete the illustration\cite{Chang:2009zb} a simplified form of the effective interaction in Ref.\,\raisebox{-1.35ex}{\mbox{\Large \cite{Maris:1997tm}}} was%
\begin{center}
\centerline{\includegraphics[clip,width=0.43\textwidth]{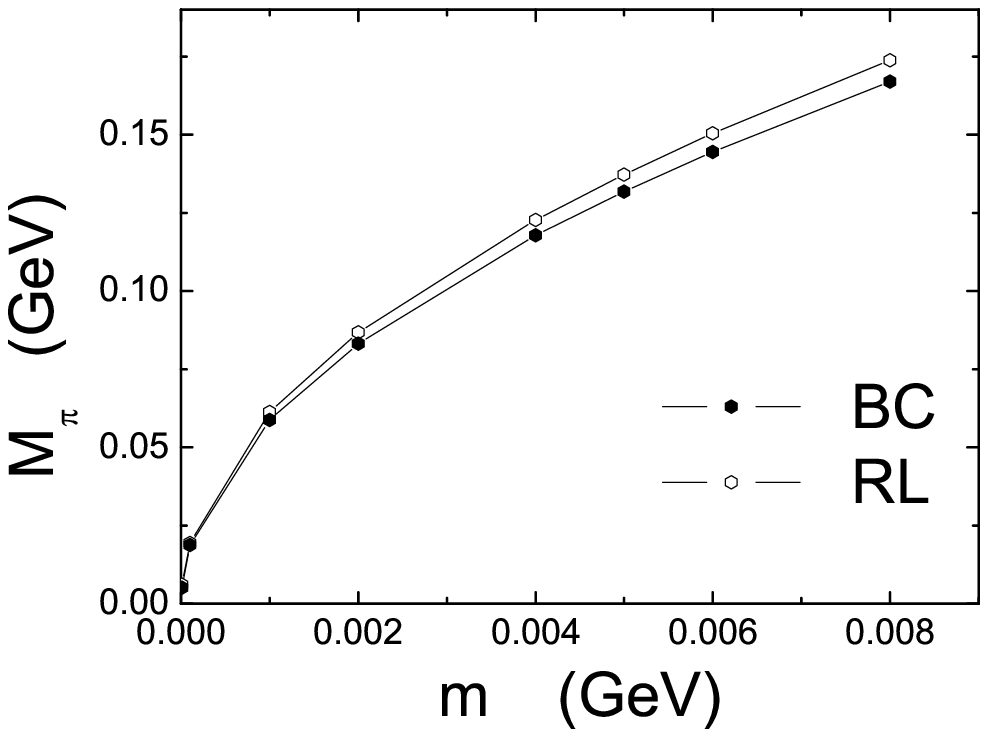}}
\vspace*{-5ex}

\centerline{\includegraphics[clip,width=0.43\textwidth]{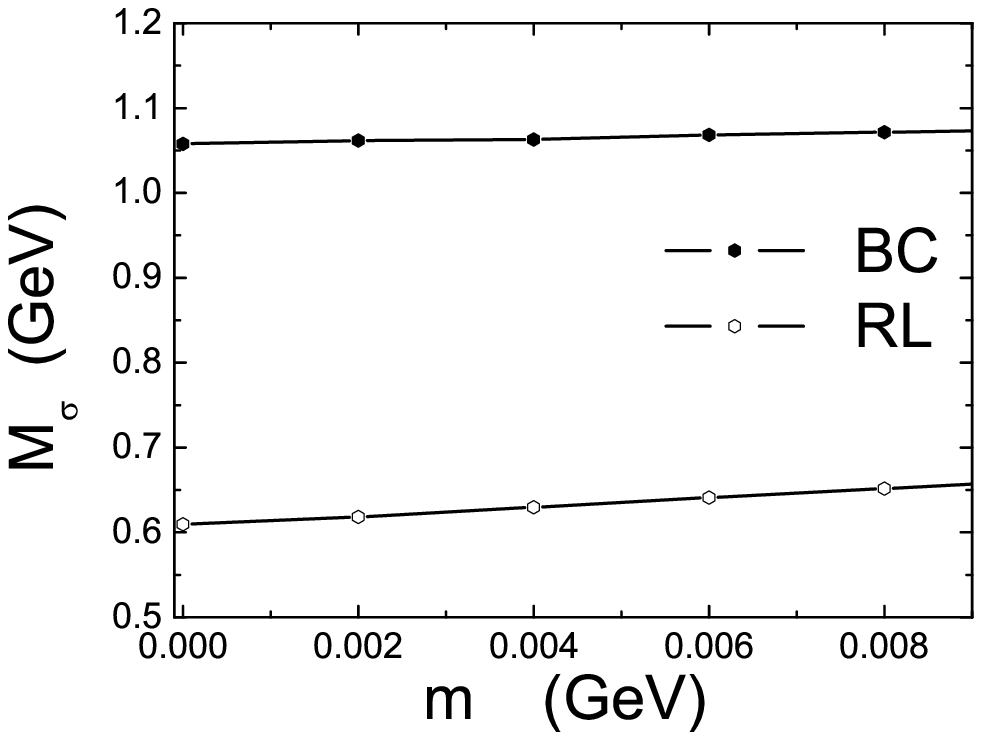}}
\vspace*{-2ex}

\figcaption{\label{massDlarge} Current-quark-mass dependence of pseudoscalar (upper panel) and scalar (lower) meson masses.  The Ball-Chiu vertex result is compared with the rainbow-ladder result.}
\end{center}
employed and two vertex \emph{Ans\"atze} were compared; viz., the bare vertex $\Gamma_\mu^g = \gamma_\mu$, which defines the rainbow-ladder truncation of the DSEs and omits vertex dressing; and the Ball-Chiu vertex,\cite{Ball:1980ay} which nonperturbatively incorporates vertex dressing associated with DCSB. 

The procedure outlined herein enables one to calculate the current-quark-mass-dependence of meson masses using a symmetry-preserving DSE truncation whose diagrammatic content is unknown.  That dependence is depicted in Fig.\,\ref{massDlarge} and compared with the rainbow-ladder result.  The $m$-dependence of the pseudoscalar meson's mass provides numerical confirmation of the algebraic fact that the axial-vector Ward-Takahashi identity is preserved by both the rainbow-ladder truncation and the BC-consistent \emph{Ansatz} for the Bethe-Salpeter kernel.  The figure also shows that the axial-vector Ward-Takahashi identity and DCSB conspire to shield the pion's mass from material variation in response to dressing the quark-gluon vertex.\cite{Roberts:2007jh,Bhagwat:2004hn}

As noted in Ref.\,\raisebox{-1.35ex}{\mbox{\Large \cite{Chang:2009zb}}}, a rainbow-ladder kernel with realistic interaction strength yields
\begin{equation}
\label{epsilonRL}
\varepsilon_\sigma^{\rm RL} := \left.\frac{2 M(0) - m_\sigma }{2 M(0)}\right|_{\rm RL} = (0.3 \pm 0.1)\,,
\end{equation} 
which can be contrasted with the value obtained using the BC-consistent Bethe-Salpeter kernel; viz., 
\begin{equation}
\label{epsilonBC}
\varepsilon_\sigma^{\rm BC} \lesssim 0.1\,.
\end{equation}
Plainly, significant additional repulsion is present in the BC-consistent truncation of the scalar BSE.

Scalar mesons are commonly identified as $^3\!P_0$ states.  This assignment reflects a constituent-quark model perspective, from which a $J^{PC}=0^{++}$ fermion-antifermion bound-state must have the constituents' spins aligned and one unit of constituent orbital angular momentum.  From this viewpoint a scalar is a spin and orbital excitation of a pseudoscalar meson.  We note that although the constituent-quark model cannot be connected with QCD, the presence of orbital angular momentum in a hadron's rest frame is a necessary consequence of Poincar\'e covariance and the vector-boson-exchange character of that theory.\cite{Bhagwat:2006pu,Bhagwat:2006xi,Cloet:2007pi} 

Extant studies of realistic corrections to the rainbow-ladder truncation show that they reduce hyperfine splitting.\cite{Bhagwat:2004hn}  Hence, with the comparison between Eqs.\,(\ref{epsilonRL}) and (\ref{epsilonBC}) one has a clear indication that in a Poincar\'e covariant treatment the BC-consistent truncation magnifies spin-orbit splitting.  This may be attributed to the influence of the quark's dynamically-enhanced scalar self-energy\cite{Roberts:2007ji} in the Bethe-Salpeter kernel.  

This feature may reasonably be expected to have a material impact on mesons with mass greater than 1\,GeV.  Indeed, \emph{prima facie} it can plausibly overcome a longstanding shortcoming of the rainbow-ladder truncation; viz., that the splitting between vector and axial-vector mesons is too small.\cite{Maris:2006ea}
This expectation is supported by Ref.\,\raisebox{-1.35ex}{\mbox{\Large \cite{Bloch:1999vka}}} wherein, using a separable Ansatz for the Bethe-Salpeter kernel which depends explicitly on the strength of DCSB, a vector--axial-vector mass-splitting is obtained that is commensurate with experiment.

\section{Baryons}
\subsection{Faddeev equation}
While a symmetry-preserving description of mesons is essential, it is only part of the story that nonperturbative QCD has to tell.  An explanation of the spectrum of baryons and the nature of interactions between them is basic to understanding the Standard Model.  The present and future resonance programmes at JLab and the Excited Baryon Analysis Center are critical elements in this effort.  They are a vital complement to the Hall-D meson programme.

QCD confines light-quarks in particle-antiparticle pairs and also in three-particle composites.  No approach to nonperturbative QCD is comprehensive if it cannot provide a unified explanation of both.  DCSB, a keystone of the Standard Model and evident in the momentum-dependence of the dressed-quark mass function -- Fig.\,\ref{gluoncloud}, is just as important to baryons as it is to mesons.  The DSEs furnish the only extant framework that can simultaneously connect both meson and baryon observables with this basic feature of QCD, having provided, e.g., a direct correlation of meson and baryon properties via a single interaction kernel, which preserves QCD's one-loop renormalisation group behaviour and can systematically be improved.\cite{Eichmann:2008ae,Eichmann:2008ef} 

In quantum field theory a baryon appears as a pole in a six-point quark Green function.  The residue is proportional to the baryon's Faddeev amplitude, which is obtained from a Poincar\'e covariant Faddeev equation that sums all possible exchanges and interactions that can take place between three dressed-quarks.  A tractable Faddeev equation for baryons\cite{Cahill:1988dx} is founded on the observation that an interaction which describes colour-singlet mesons also generates nonpointlike quark-quark (diquark) correlations in the colour-$\bar 3$ (antitriplet) channel.\cite{Cahill:1987qr}  The lightest diquark correlations appear in the $J^P=0^+,1^+$ channels\cite{Burden:1996nh,Maris:2002yu} and hence today only they are retained in approximating the quark-quark scattering matrix that appears as part of the Faddeev equation.\cite{Eichmann:2008ef,Cloet:2008re}  

While diquarks do not appear in the strong interaction spectrum,\cite{Bender:1996bb,Bhagwat:2004hn} the attraction between quarks in this channel justifies a picture of baryons in which two quarks are always correlated as a colour-$\bar 3$ diquark pseudoparticle, and binding is effected by the iterated exchange of roles between the bystander and diquark-participant quarks.   Here it is important to emphasise strongly that QCD supports \emph{nonpointlike} diquark correlations.\cite{Maris:2004bp,Alexandrou:2006cq}  Models that employ pointlike diquark degrees of freedom cannot be connected with QCD.  This, however, is a defect they share with all approaches that employ pointlike-constituent degrees of freedom.  It is therefore not surprising that experimental observations contradict the spectroscopic predictions of such models; e.g., in connection with the so-called \emph{missing resonance} problem, the best information available today indicates that even some listed $\ast\ast\ast\ast$-resonances should be discarded.\cite{tshlee} 

\subsection{Strangeness sigma-term}
Numerous properties of the nucleon have been computed using the Faddeev equation just described.  An example is the nucleon's $\sigma$-term, with the result:\cite{Flambaum:2005kc}
\begin{equation}
\label{sigmaN}
f_N^u:=\frac{\sigma_N}{M_N} \approx\,6\%.
\end{equation}
This measures the contribution to the nucleon's mass from the explicit chiral symmetry breaking term associated with $u$- and $d$-quarks in QCD's Lagrangian.  Of material additional interest is the contribution to the nucleon's mass from the $s$-quark mass term.  

We have estimated this by analysing the dressed-quark $\sigma$-term\cite{Flambaum:2005kc,Holl:2005st} using a gap equation that incorporates $\pi$- and $K$-loop contributions and which has previously been used to estimate the strangeness contribution to the nucleon's magnetic moment: $\mu_p^S\approx -0.02\,$nuclear magnetons.\cite{Cloet:2008fw}  The model yields
\begin{equation}
\label{sigmaQ}
f_U^u:= \frac{m_u}{M_u} \frac{d M_u}{d m_u} = 6.5\% , \quad f_U^s:=\frac{m_s}{M_u} \frac{d M_u}{d m_s} = 2.4\%.
\end{equation}
Comparing Eqs.\,(\ref{sigmaN}) and (\ref{sigmaQ}), one observes $f_N^u \simeq f_U^u$.  One would have equality between these two quantities in a weak-binding independent particle model.  We therefore anticipate that
\begin{equation}
\label{fNs}
f_N^s \approx f_U^s = 2.4\%.
\end{equation}
The results in Eqs.\,(\ref{sigmaN}) and (\ref{fNs}) agree with those inferred recently\cite{Young:2009zb} from numerical simulations of lattice-regularised QCD.
 
We observe that pseudoscalar meson exchange is attractive between a fermion and antifermion.  Hence one knows with certainty \emph{a priori} that a valid calculation of pseudoscalar-meson-loop contributions to the gap and Bethe-Salpeter equations must show a reduction in both the mass of the fermion the mesons dress and the bound-state they bind.  One check on a study is the effect of the loops on the quark condensate: in a realistic truncation they must reduce its size, as is found, e.g., in the model employed above.  It follows from this fact and the axial-vector Ward-Takahashi that in the neighbourhood of the chiral limit
\begin{equation}
\left. f_\pi^2 m_\pi^2 \right|_{\rm with~loops} < 
\left. f_\pi^2 m_\pi^2\right|_{\rm without~loops} \,.
\end{equation}
The symmetry-preserving inclusion of pion-loop corrections to the gap and bound-state equations is challenging but progress has been made.\cite{Fischer:2007ze}  
In a recent application of this proposed method\cite{Fischer:2008wy} some results were obtained that conflict with our statements.  Their origin is now understood (a confusion in the renormalisation prescription) and they will be corrected in a forthcoming article.\cite{CFRWprivate}
%

\subsection{Neutron electromagnetic form factors} 
A comprehensive analysis of nucleon electromagnetic form factors using the DSEs is available\cite{Cloet:2008re} and%
\begin{center}
\includegraphics[clip,width=0.46\textwidth]{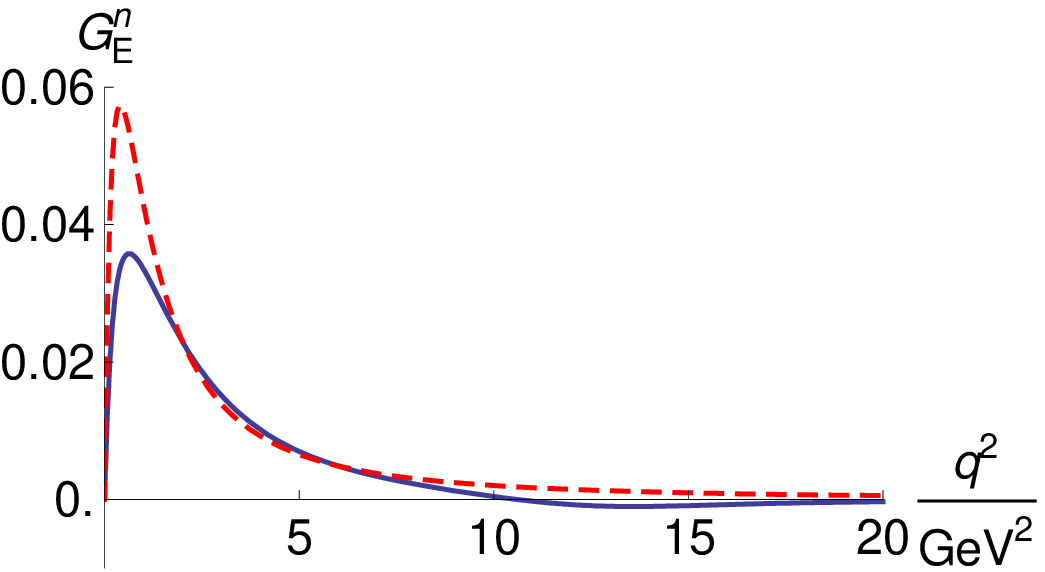}
\figcaption{\label{GEneutron} 
Sachs neutron electric form factor: \emph{solid curve} -- DSE prediction;\protect\cite{Cloet:2008re}
\emph{dashed curve} -- a 2004 parametrisation of data.\protect\cite{Kelly:2004hm}  New JLab Hall-A data on the neutron form factor at $Q^2 = 1.71,2.51,3.47\,$GeV$^2$ will soon be available.\protect\cite{bogdan}}
\end{center}
\begin{center}
\includegraphics[clip,width=0.46\textwidth]{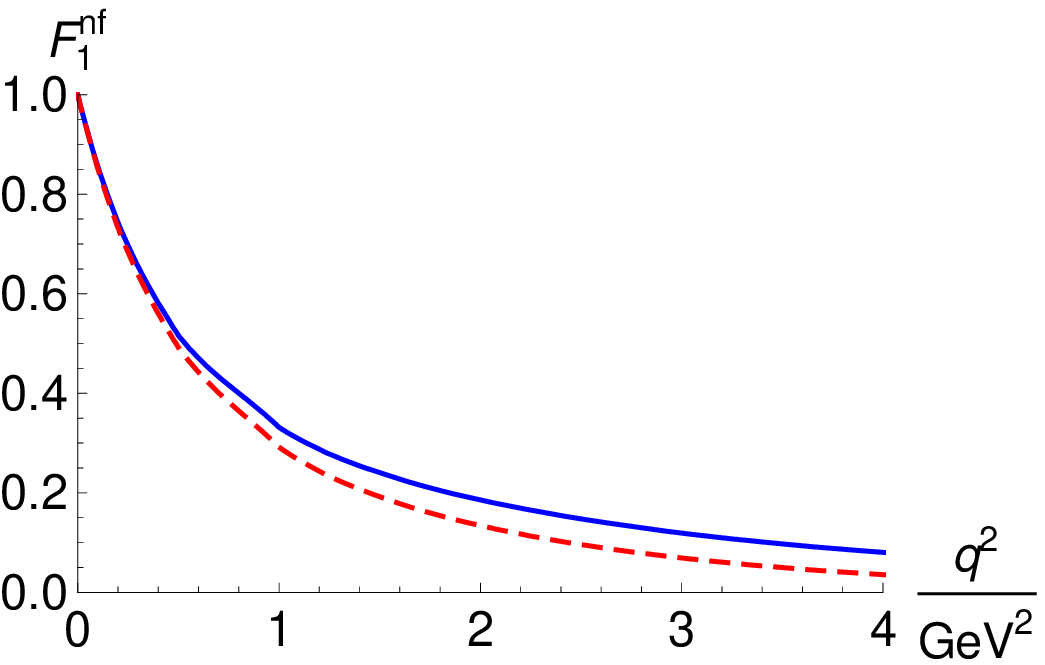}
\includegraphics[clip,width=0.46\textwidth]{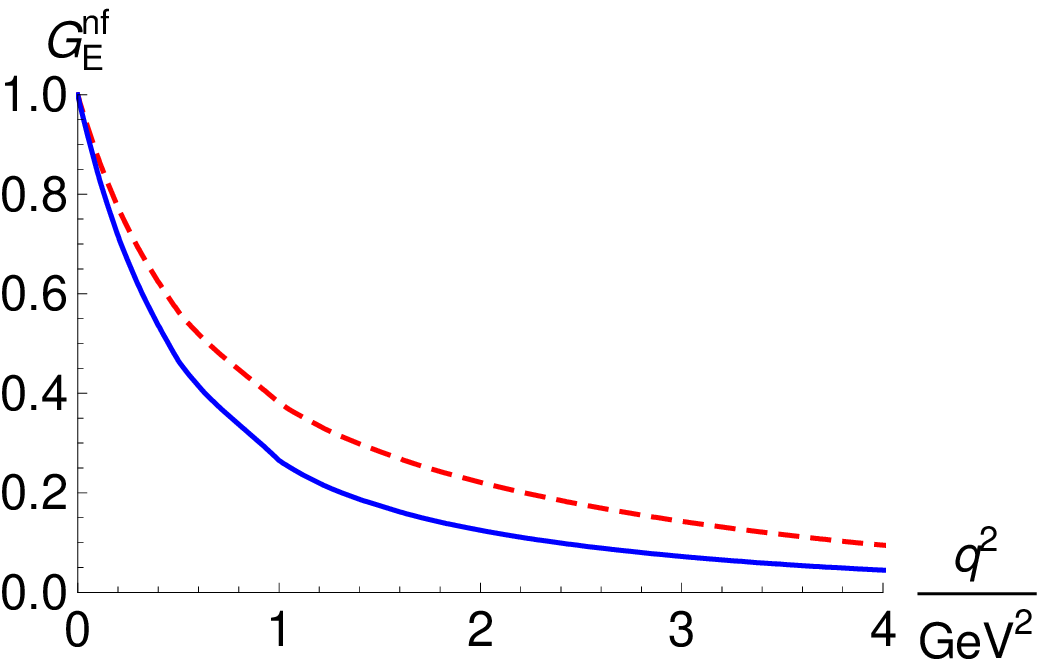}
\figcaption{\label{GEneutronud} 
Flavour decomposition of neutron form factors:\protect\cite{Cloet:2008re} \emph{Upper panel} -- Dirac; \emph{Lower panel} -- Sachs electric.  In both panels: \emph{solid curve} -- $f$=$d$-quark; \emph{dashed curve} -- $f$=$u$-quark.  All form factors normalised to unity at $q^2=0$.  In reality: 
$F_1^{nd}(0)=G_E^{nd}(0)=-\frac{2}{3}$ and $F_1^{nu}(0)=G_E^{nu}(0)=\frac{2}{3}$; $F_1^{nd}(q^2)$ is negative definite, $G_E^{nd}(q^2)$ becomes positive at $q^2\approx 9\,$GeV$^2$, $F_1^{nu}(q^2)$ becomes negative at $q^2\approx 7\,$GeV$^2$ and $G_E^{nu}(q^2)$ becomes negative at $q^2\approx 10\,$GeV$^2$.}
\end{center}
the calculation of nucleon-to-resonance transition form factors is underway.  

These studies compute a dressed-quark core contribution to the form factors, which is defined by the%
\begin{center}
\includegraphics[clip,width=0.30\textwidth]{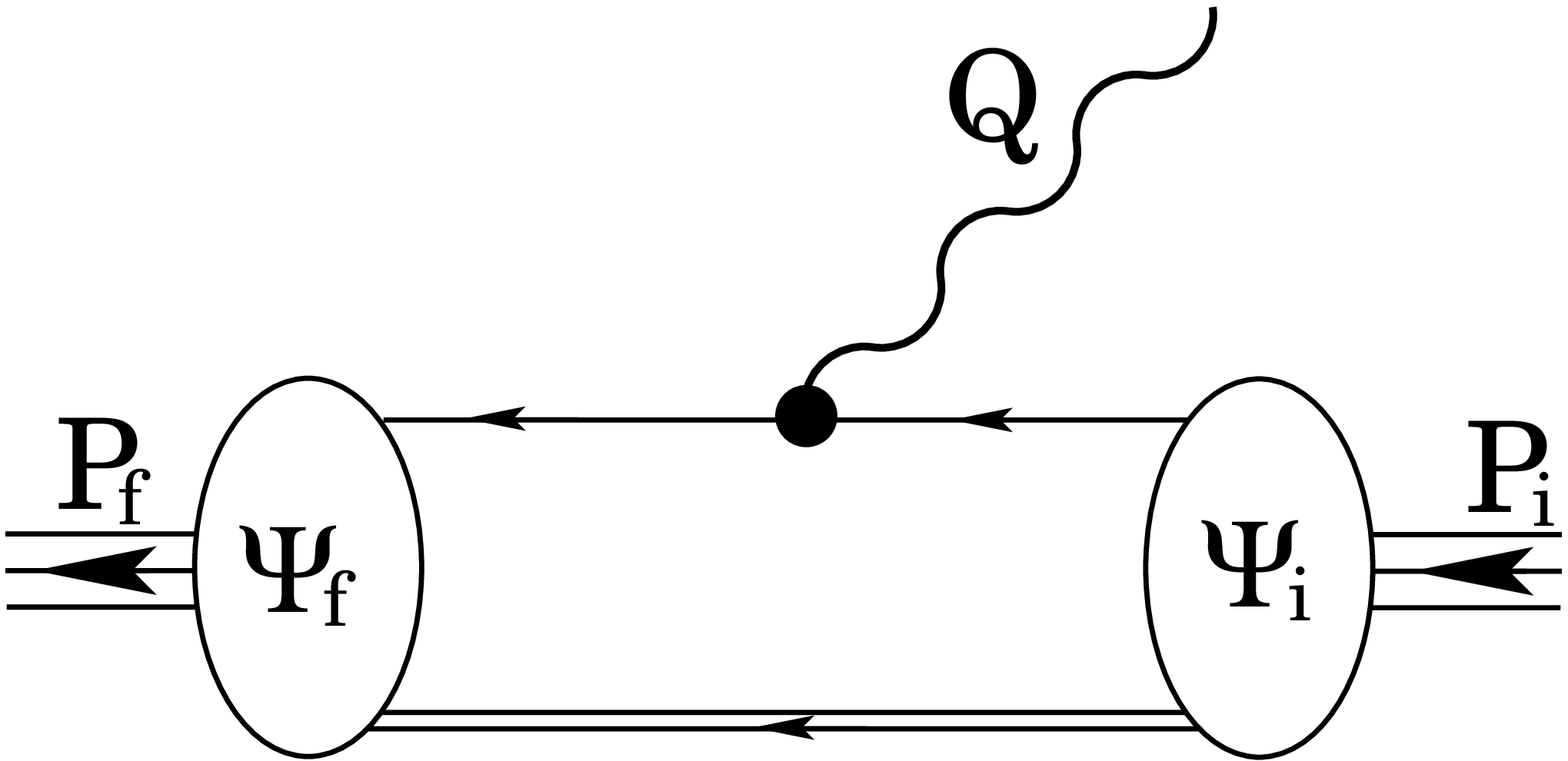}
\figcaption{\label{impulse} Bystander quark vertex, one of six diagrams that contribute to a conserved current for on-shell nucleons described by the Faddeev equation solution, $\Psi_{i,f}$.\protect\cite{Oettel:1999gc}  The single line represents $S(p)$, the dressed-quark propagator, and the double line is the diquark propagator.}
\end{center}
solution of the Poincar\'e covariant Faddeev equation.  A particular feature of the study cited is a separation of form factor contributions into those from different
diagram types and correlation sectors, and subsequently a flavour separation for each of these.  Amongst the extensive body of results that one might highlight are: both the neutron (Fig.\,\ref{GEneutron}) and proton Sachs electric form factor possess a zero; and, owing to the presence of axial-vector quark-quark correlations, $r_1^{n,u}>r_1^{n,d}$ but $r_E^{n,u}<r_E^{n,d}$ (Fig.\,\ref{GEneutronud}).

In vanishing twice on the domain accessible to reliable calculation, at the origin owing to charge neutrality, and at $Q^2\approx 11\,$GeV$^2$ owing to dynamics, the
neutron's electric form factor is special.  The origin of this second zero can be explained and is consistent with intuition.\cite{Cloet:2008re}  For example, consider $G_E^{n,q_b}$, which is the contribution to the form factor from a \emph{bystander quark}; viz., the contribution from a process in which the photon strikes a quark that is neither within a diquark nor breaking away from one, illustrated in Fig.\,\ref{impulse}.  (NB.\ This is one contribution to the quantity plotted in Fig.\,\ref{GEneutronud}, which is the total $f$-quark contribution to the form factor.)  $G_E^{n,q_b}$ is negative at small-$Q^2$ because the scalar diquark component of the Faddeev amplitude is dominant and that is paired with a $d$-quark bystander in the neutron.  This dressed-quark is responsible for the preponderance of negative charge at long range.  $G_E^{n,q_b}$ is positive at large $Q^2$ because $F_2^n$ dominates on that domain, which focuses attention on the axial-vector diquark component of the Faddeev amplitude.  The positively charged $u$-quark is most likely the bystander quark in these circumstances.  


\subsection{Neutron charge distribution}
These features are manifest in the configuration-space charge density, obtained through a three-dimensional Fourier-Transform; namely, 
\begin{equation}
\label{rhonr}
\rho_n(r) = \frac{1}{2\pi^2 r} \int_0^\infty \! dq\, q \,\sin (q r)\,  G_E^n(q^2)\,,
\end{equation}
\begin{center}
\includegraphics[clip,width=0.46\textwidth]{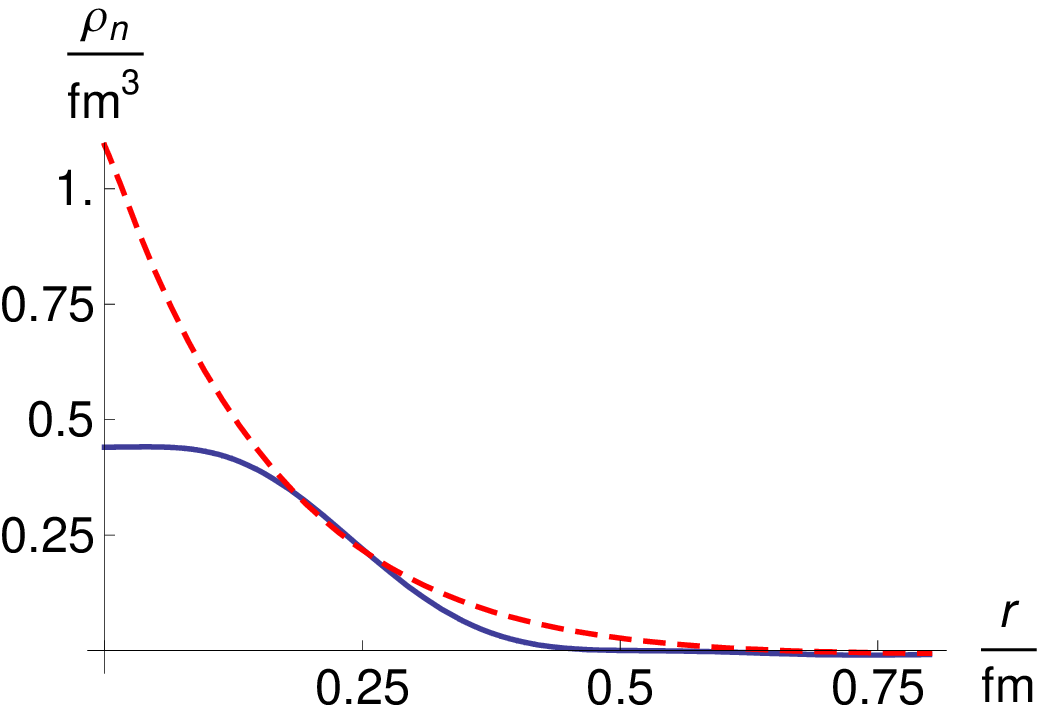}
\includegraphics[clip,width=0.46\textwidth]{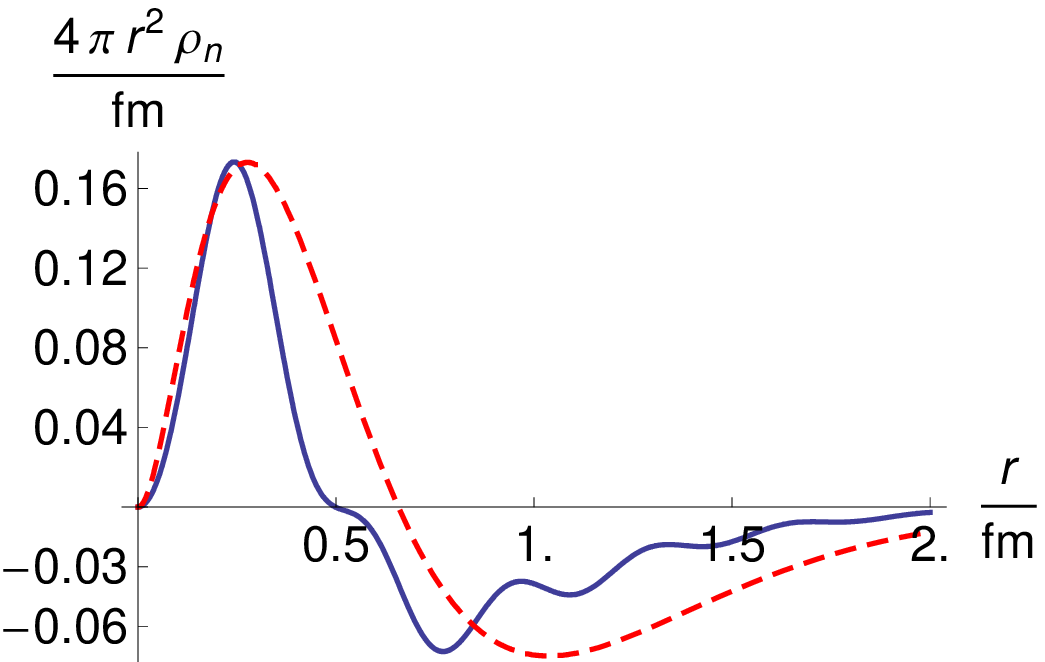}
\figcaption{\label{rhoneutron} 
Distribution of charge within the neutron evaluated from the three-dimensional Fourier transform in Eq.\,(\protect\ref{rhonr}): \emph{solid curve} -- DSE prediction;\protect\cite{Cloet:2008re}
\emph{dashed curve} -- a 2004 parametrisation of data.\protect\cite{Kelly:2004hm}
}
\end{center}
which is depicted in Fig.\,\ref{rhoneutron}.  To compute $\rho_n(r)$ we inferred $G_E^n(q^2)$ from the calculated ratio $\mu_n G_E^n(q^2)/G_M^n(q^2)$ (Fig.\,16, Ref.\,\raisebox{-1.35ex}{\mbox{\Large \cite{Cloet:2008re}}}) multiplied by
the empirical dipole: $1/[1+q^2/(0.84\,{\rm GeV})^2]^2$.  This procedure corrects for the deliberate omission of pion cloud effects in Ref.\,\raisebox{-1.35ex}{\mbox{\Large \cite{Cloet:2008re}}}. The result is depicted in Fig.\,\ref{GEneutron}.  Caveats on the interpretation of $\rho_n(r)$ as a quantum mechanical charge density are discussed in Sec.\,4 of Ref.\,\raisebox{-1.35ex}{\mbox{\Large \cite{Cloet:2008re}}}.  These observations notwithstanding, the mapping between the $q^2$-dependence of the Sachs electric form factor and the charge density is intuitively appealing and instructive. 

In the comparison made in Fig.\,\ref{rhoneutron} between the DSE prediction\cite{Cloet:2008re} and a 2004 parametrisation of data,\cite{Kelly:2004hm} two features are striking: a significant depletion of positive charge at the core of the neutron accompanied by an increased concentration of negative charge toward the surface; and oscillations in the charge distribution.  NB.\ We computed the charge density arising only from the dressed-quark core.

The depletion of charge is associated with the second zero in $G_E^n(q^2)$ and the domain of negative support which follows.  The amount of charge depletion is determined by the magnitude of $G_E^n(q^2)$ at it's min-%
\begin{center}
\includegraphics[clip,width=0.46\textwidth]{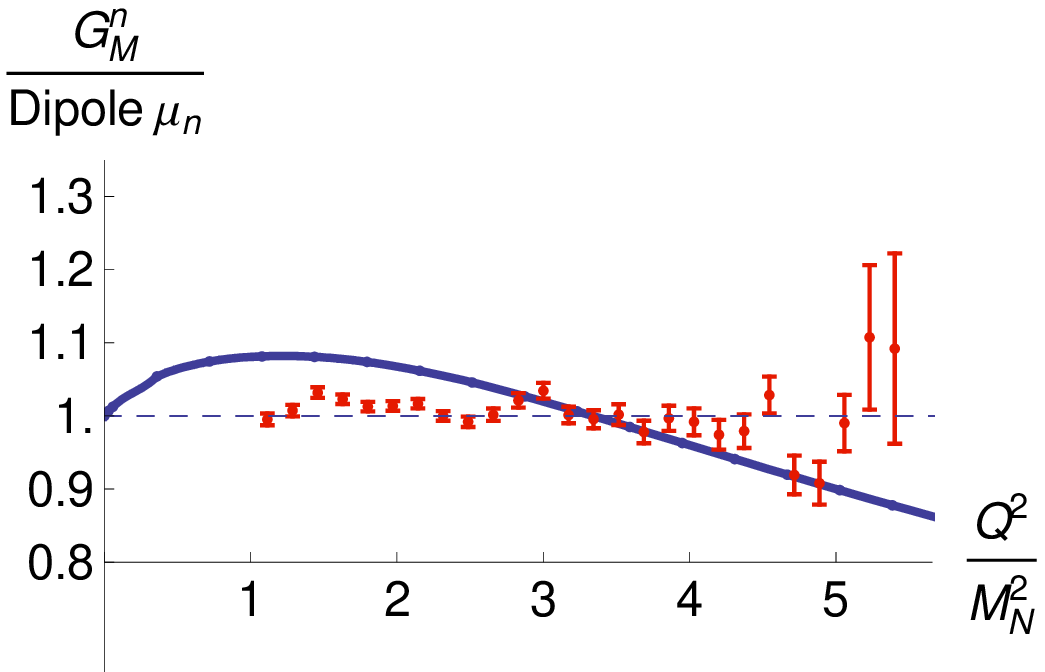}
\figcaption{\label{GMdipole} 
Sachs neutron magnetic form factor divided by dipole fit: \emph{solid curve} -- DSE prediction;\protect\cite{Cloet:2008re}
\emph{points} -- contemporary experiment.\protect\cite{Lachniet:2008qf}  As explained in Sec.\,8 of the DSE study,\protect\cite{Cloet:2008re} pseudoscalar meson loops, deliberately omitted in the calculation displayed here, will work to flatten the prediction, with greatest impact for $Q^2\lesssim 3 M_N^2$.}
\end{center}
imum: should the magnitude exceed a critical value, $\rho_n(r)$ would become negative in the neighbourhood of $r=0$.

The oscillations at $r\gtrsim 0.5\,$fm are connected with the shape of $G_E^n(q^2)$ on its domain of positive support.  Crucial to their appearance and nature are the following features: $G_E^n(0)=0$; the location and magnitude of the maximum of $G_E^n(q^2)$; and the fact that the domain of positive support is bounded.

Given that it is common practice to compare nucleon form factors with an empirical dipole, it is of interest to compare the DSE results with a dipole parametrisation.  To that end we fitted $1/(1+Q^2/m_D^2)^2$ to the DSE result on $2\leq Q^2/M_N^2 < 9$.  This domain excludes the region whereupon pion cloud effects are significant and maximises coverage of the domain on which the quark-core calculation is most reliable.  The fit produced $m_D= 1.06\, M_N$, which is just 18\% larger than the empirical dipole mass, $m_D^{\rm emp}=0.89\,M_N$.  The ratio obtained with the computed dipole mass is depicted in Fig.\,\ref{GMdipole} and compared with a modern experimental determination.\cite{Lachniet:2008qf}

\section{Perspective}
Plainly, much has been learnt from the application of Dyson-Schwinger equations (DSEs) to problems in nonperturbative QCD.  This process will continue.  

For example, comparison between DSE results and forthcoming precision data on nucleon form factors, both elastic and resonance-transition, holds promise as a means by which to chart the momentum evolution of the dressed-quark mass function and therefrom the infrared behavior of QCD's $\beta$-function.  In particular, it should enable the unambiguous location of the transition boundary between the constituent- and current-quark domains that is signalled by the sharp drop apparent in Fig.\,\ref{gluoncloud} and which can likely be related to an infrared inflexion point in QCD's running coupling, whose properties are determined by the $\beta$-function.

Contemporary theory indicates that this transition boundary lies at $p^2 \sim 0.6\,$GeV$^2$.  Since a probe's input momentum $q$ is principally shared equally amongst the dressed-quarks in elastic and transition processes, then each can be considered as absorbing a momentum fraction $q/3$.  Thus in order to scan the behaviour of the mass function on the domain $p^2\in [0.5,1.0]\,$GeV$^2$ one requires $q^2\in [5,10]\,$GeV$^2$.  This domain will become accessible after completion of the upgrade underway currently at JLab.

\medskip

\acknowledgments{CDR is grateful to the organisers of NSTAR2009 for arranging a very rewarding meeting.  We acknowledge valuable communications with V.~Burkert, G.\,P.~Gilfoyle, R.~Gothe, T.-S.\,H.~Lee, Y.\,X.~Liu, V.~Mokeev, R.\,A.~Schumacher, B.~Wojtsekhowski, R.\,D.~Young and H.\,S.~Zong.}

\end{multicols}

\vspace{-2mm}
\centerline{\rule{80mm}{0.1pt}}
\vspace{2mm}

\begin{multicols}{2}

\newcounter{dumbone}
\setcounter{dumbone}{0}

\end{multicols}

\end{document}